\documentclass[twocolumn,showpacs,preprintnumbers,amsmath,amssymb,superscriptaddress]{revtex4}

\usepackage{graphicx}
\usepackage{dcolumn}
\usepackage{bm}

\newcommand{\bra}[1]{\langle #1|}
\newcommand{\ket}[1]{|#1\rangle}

\begin{document}

\title{Tunnel magnetoresistance of quantum dots coupled to ferromagnetic
leads \\ in the sequential and cotunneling regimes}

\author{Ireneusz Weymann}
\affiliation{Department of Physics, Adam Mickiewicz University,
61-614 Pozna\'n, Poland} \affiliation{Institut f\"ur Theoretische
Festk\"orperphysik, Universit\"at Karlsruhe, 76128 Karlsruhe,
Germany}

\author{J\"urgen K\"onig}
\affiliation{Institut f\"ur Theoretische Physik III,
Ruhr-Universit\"at Bochum, 44780 Bochum, Germany}

\author{Jan Martinek}
\affiliation{Institut f\"ur Theoretische Festk\"orperphysik,
Universit\"at Karlsruhe, 76128 Karlsruhe, Germany}
\affiliation{Institute of Molecular Physics, Polish Academy of
Sciences, 60-179 Pozna\'n, Poland}\affiliation{Institute for
Materials Research, Tohoku University, Sendai 980-8577, Japan}

\author{J\'ozef Barna\'s}
\affiliation{Department of Physics, Adam Mickiewicz University,
61-614 Pozna\'n, Poland} \affiliation{Institute of Molecular
Physics, Polish Academy of Sciences, 60-179 Pozna\'n, Poland}

\author{Gerd Sch\"on}
\affiliation{Institut f\"ur Theoretische Festk\"orperphysik,
Universit\"at Karlsruhe, 76128 Karlsruhe, Germany}

\date{\today}

\begin{abstract}
We study electronic transport through quantum dots weakly coupled
to ferromagnetic leads with collinear magnetization directions.
Tunneling contributions of first and second order in the
tunnel-coupling strength are taken into account. We analyze the
tunnel magnetoresistance (TMR) for all combinations of linear and
nonlinear response, at or off resonance, with an even or odd
dot-electron number. Different mechanisms for transport and spin
accumulation the various regimes give rise to different TMR
behavior.

\end{abstract}

\pacs{72.25.Mk, 73.63.Kv, 85.75.-d, 73.23.Hk}

\maketitle

\section{Introduction}

The study of spin-polarized electron transport through
nanostructures with strong Coulomb interaction is a relatively new
field of theoretical and experimental research, residing in the
intersection of the fields of spintronics
\cite{wolf01,gregg02,loss02,maekawa02} and transport through
nanostructures \cite{averin,nato,schon}, respectively. The
interplay of finite spin polarization and Coulomb blockade gives
rise to a complex transport behavior in which both the electrons'
charge and spin degree of freedom play a role \cite{barnas98}. A
convenient minimal model system to study this interplay consists
of a single-level quantum dot coupled through tunnel barriers to
ferromagnetic electrodes. Experimentally such systems may be
realized in various ways, including self-assembled dots in
ferromagnetic semiconductors \cite{chye}, ultrasmall aluminum
nanoparticles \cite{ralph02}, carbon nanotubes
\cite{tsukagoshi,zhao,nygard}, or single molecules
\cite{pasupathy}.

The properties of spin-polarized transport through single magnetic
tunnel junctions have already proven technological relevance in
information-storage devices based on the tunnel magnetoresistance
(TMR) effect, i.e., the observation that the current flowing
through the junction depends on the relative orientation of the
leads' magnetizations. It is maximal for the parallel and minimal
for the antiparallel configuration. Quantitatively, it can be
characterized by
\begin{equation}
  {\rm TMR} = \frac{I_{\rm P}-I_{\rm AP}}{I_{\rm AP}} \,
\end{equation}
where $I_{\rm P}$ and $I_{\rm AP}$ are the currents for the
parallel and antiparallel configuration, respectively. Julliere
found \cite{julliere} that the TMR for a single tunnel junction is
related to the degree $p$ of spin polarization of the leads'
density of states, $p = (\rho^+ - \rho^-)/(\rho^+ + \rho^-)$, by
$\text{TMR}^{\rm Jull}=2p^2/(1-p^2)$, where $\rho^{+}$ and
$\rho^{-}$ are the spin-majority and spin-minority densities of
states in the electrodes, respectively. Julliere's formula
immediately follows from the fact that the transmission
probability of an electron with spin $\sigma$ through the barrier
is proportional to the product of the (spin-dependent) densities
of states for spin $\sigma$ in source and drain.

Once a nanoscopic island is placed in between the ferromagnetic
leads the situation becomes much more complex for two reasons.
First, there are different types of transport processes that
depend on the leads' spin polarization in a different manner, such
as sequential tunneling, non-spin-flip, and spin-flip cotunneling
(for non-spin-flip cotunneling an electron of given spin is
transferred through the system, while for spin-flip cotunneling
both the spin of the transferred electron as well as the dot spin
changes during the process). Second, a non-equilibrium spin
accumulation can partially polarize the island, which, in turn,
affects the total transmission through the device. Therefore, the
TMR will, in general, deviate from Julliere's value. It will,
furthermore, be different for different transport regimes. The
measurement of the TMR as a function of temperature, bias and gate
voltages, will, thus, reveal information about the underlying
transport processes as well as the spin accumulation on the
island.

Spin-dependent transport through a single-level quantum dot in the
sequential-tunneling regime with collinearly magnetized leads has
been analyzed in Refs.~\onlinecite{bulka,rudzinski,cottet}. This
has been extended \cite{koenigPRL03,braun,rudzinski04,braig04} to
noncollinear configurations with arbitrary relative angle, for
which a precession of the dot spin about an intrinsic exchange
field gives rise to non-trivial dependence of the angle-dependent
conductance. In the present paper, we analyze the TMR for
collinear magnetization beyond sequential tunneling. This covers
the Coulomb-blockade regime, in which sequential tunneling is
exponentially suppressed, and transport is dominated by
cotunneling
\cite{odintsov,nazarov,kang,franceschi01,kogan04,zumbuhl,ciorga}.
But even when sequential tunneling is possible, second-order
corrections to the current become important for increasing
tunnel-coupling strengths. This includes the above-mentioned
cotunneling processes but also terms associated with
renormalization of level position and tunnel-coupling strength
\cite{koenigdiss}. Recently, we studied spin-dependent transport
for a specific transport regime, namely, cotunneling deep inside
the Coulomb-blockade valley \cite{weymann04}.

Our objective for the present paper is to analyze the TMR in the
full parameter space defined by the gate and bias voltages. This
includes the linear- and nonlinear-response regime as well as the
cases of even and odd dot occupation. We find that the TMR reaches
Julliere's value only when the transport is fully carried by
non-spin-flip cotunneling. This happens in the Coulomb-blockade
valleys in which the dot is either empty or doubly occupied, where
the dot remains unpolarized, as well as for large bias voltage in
the Coulomb-blockade valley with an odd dot-electron number. For
all other regimes, though, the TMR is reduced below Julliere's
value.

\section{Model}

\begin{figure}[t]
  \includegraphics[width=0.9\columnwidth]{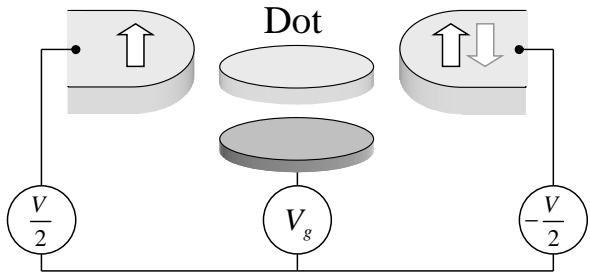}
  \caption{\label{qd}Single-level quantum dot coupled to ferromagnetic
    leads. The magnetic moments of the electrodes are either parallel or
    antiparallel to each other.}
\end{figure}

We consider transport through a single-level quantum dot. The dot
is coupled to two ferromagnetic electrodes with collinear, i.e.,
either parallel or antiparallel, magnetizations, see
Fig.~\ref{qd}. The dot level $\varepsilon$ can be tuned by a gate
voltage, but is independent of the symmetrically-applied transport
voltage.

We model the system by an Anderson-like Hamiltonian of the form
\begin{equation}
  H=H_{\rm L}+H_{\rm R}+ H_{\rm D}+H_{\rm T}.
\end{equation}
The first and second terms represent the left and right reservoirs
of noninteracting electrons, $H_{r}=\sum_{q\sigma} \varepsilon_{r
q\sigma}c^{\dagger}_{r q\sigma} c_{r q\sigma}$, for $r ={\rm
L,R}$, where $c^{\dagger}_{r q\sigma}$ ($c_{r q\sigma}$) is the
creation (annihilation) operator of an electron with wave number
$q$ and spin $\sigma$ in the lead $r$, whereas $\varepsilon_{r
q\sigma}$ denotes the corresponding single-particle energy. The
dot is represented by
\begin{equation}
  H_{\rm D}=\sum_{\sigma=\uparrow,\downarrow}
  \varepsilon d^{\dagger}_{\sigma} d_{\sigma}+U
  d^{\dagger}_{\uparrow} d_{\uparrow}
  d^{\dagger}_{\downarrow}d_{\downarrow},
\end{equation}
with $d^{\dagger}_{\sigma}$ ($d_{\sigma}$) creating (annihilating)
an electron on the dot with spin $\sigma$ and energy
$\varepsilon$, and $U$ is the charging energy for double
occupancy. There are four possible states for the quantum dot:
empty dot ($\chi =0$), singly-occupied dot with a spin-up ($\chi
=\;\uparrow$) or spin-down ($\chi =\;\downarrow$) electron, and
doubly-occupied dot ($\chi={\rm d}$). Tunneling between dot and
leads is described by
\begin{equation}
  H_{\rm T}=\sum_{r={\rm L,R}}\sum_{q\sigma}\left( t_{r
  q\sigma}c^{\dagger}_{r q\sigma} d_{\sigma}+t^{*}_{r
  q\sigma}d_{\sigma}^{\dagger}c_{r q\sigma}\right) \, ,
\end{equation}
where $t_{r q\sigma}$ are the tunnel matrix elements. Tunneling
gives rise to an intrinsic broadening $\Gamma^{\sigma}$ of the dot
levels, given by the Fermi-golden-rule expression
$\Gamma^{\sigma}= \sum_{r={\rm L,R}}\Gamma^{\sigma}_r$, with
$\Gamma^{\sigma}_r= 2\pi \sum_{q} |t_{r q\sigma}|^2
\delta(\omega-\varepsilon_{r q\sigma})$. Assuming the matrix
elements $t_{r q\sigma}$ to be independent of the wave number and
spin orientation, we get $\Gamma_{r}^{\sigma}= 2\pi |t_{r}|^2
\rho_{r}^{\sigma}$, with $\rho_{r}^{\sigma}$ denoting the
spin-dependent density of states in lead $r$. In the following we
assume the latter to be independent of energy within the electron
band. Furthermore, we introduce the degree of spin polarization
$p_{r}=(\rho_{r}^{+}- \rho_{r}^{-})/ (\rho_{r}^{+}+ \rho_{r}^{-})$
of lead $r$, and express the four respective couplings in terms of
spin polarization as $\Gamma_{r}^{+(-)}=\Gamma_{r}(1\pm p_{r})$,
where $\Gamma_{r}= (\Gamma_{r}^{+} +\Gamma_{r}^{-})/2$. In
general, the leads may have different spin polarizations and/or
coupling strengths to the dot. In the following, however, we
assume $p_{\rm L} = p_{\rm R} \equiv p$ and $\Gamma_{\rm
L}=\Gamma_{\rm R}\equiv\Gamma/2$. In the weak coupling regime,
typical values of the dot-lead coupling strength $\Gamma$ are of
the order of tens of $\mu$eV \cite{kogan04}.

\section{Method and Transport Equations}

\begin{figure}[b]
  \includegraphics[width=1\columnwidth]{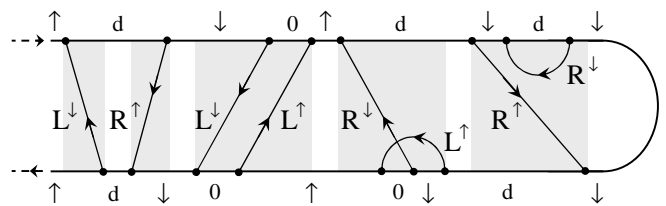}
  \caption{\label{densmat}An example for the time evolution of the
    reduced density matrix. The grey regions define irreducible diagrams
    of first and second order in tunneling, respectively.
    The direction of each tunneling line indicates whether an
    electron of respective spin leaves or enters the dot, thus,
    leading to a change of the dot state, as indicated on the forward
    and backward Keldysh propagators.}
\end{figure}

We calculate the transport properties of the system by making use
of a real-time diagrammatic technique
\cite{koenigdiss,diagrams,thielmann}. Its main idea is to
integrate out the electronic degrees of freedom in the leads in
order to arrive at an effective description of the dot subsystem.
The dynamics of the subsystem is then described by a reduced,
four-dimensional, density matrix with density matrix elements
$P_{\chi _2}^{\chi _1}(t)$. The time evolution of the reduced
system can be represented graphically as a sequence of irreducible
diagrams on the Keldysh contour. An example of such time evolution
is shown in Fig.~\ref{densmat},
where the upper and lower branches of the Keldysh contour
represent the forward and backward propagators. Tunneling is
represented by vertices, that are connected in pairs by tunnel
lines. Each grey region in Fig.~\ref{densmat} defines an
irreducible diagram that corresponds to a transition of the dot
state. First- and second-order transport in the tunnel-coupling
strength $\Gamma$ is described by diagrams containing one or two
tunnel lines, respectively. Since we consider only collinear
magnetic configurations of the leads and tunneling is spin
conserving, the natural choice of the spin-quantization axis
results in vanishing of all non-diagonal density matrix elements,
and only the diagonal ones, $P_{\chi}^{\chi}\equiv P_{\chi}$, need
to be considered. They are nothing but the probability to find the
dot in state $\chi$.

The time evolution of the reduced density matrix is governed by a
generalized master equation \cite{koenigdiss} that in the
stationary limit reduces to
\begin{equation}
\label{master}
  0=\sum_{\chi}\Sigma_{\chi^\prime \chi} P_{\chi} \, ,
\end{equation}
where $\Sigma_{\chi^\prime \chi}$ describes the irreducible
diagram parts with transitions from state $\chi$ to $\chi^\prime$.
The electric current is given by
\begin{eqnarray}
  I&=&-\frac{i e}{2\hbar}\sum_{\chi\chi^\prime}
  \Sigma_{\chi^\prime\chi}^{\text{I}} P_{\chi} \, ,
\end{eqnarray}
where the self energy $\Sigma_{\chi^\prime\chi}^{\text{I}}$ is
modified as compared to $\Sigma_{\chi^\prime\chi}$ to account for
the number of electrons transferred through the barriers. The
rules to calculate $\Sigma_{\chi^\prime\chi}$ and
$\Sigma_{\chi^\prime\chi}^{\text{I}}$ are given in the appendix.

Our goal is to calculate the current up to second order in the
tunnel-coupling strength $\Gamma$. For this, we first expand the
self-energies $\Sigma_{\chi^\prime\chi}$ and
$\Sigma_{\chi^\prime\chi}^{\text{I}}$ order by order,
\begin{equation}
  \Sigma_{\chi^\prime\chi} = \Sigma_{\chi^\prime\chi}^{(1)}+
  \Sigma_{\chi^\prime\chi}^{(2)}+\dots \; ,
\end{equation}
where the order corresponds to the number of tunnel lines of a
diagram. Consequently, the entire problem is reduced to the
calculation of all the self-energies with the aid of the
diagrammatic rules.

For an accurate perturbation expansion of the current, we also
need to expand the probabilities in orders of $\Gamma$,
\begin{equation}
  P_{\chi} = P_{\chi}^{(0)}+P_{\chi}^{(1)}+\dots \, ,
\end{equation}
with the normalization condition
\begin{equation}
  \sum_{\chi} P^{(m)}_{\chi}= \delta_{m,0} \; .
\end{equation}
The first- and second-order contributions to the current are then
given by
\begin{eqnarray}
  I^{(1)}&=&-\frac{i e}{2\hbar}\sum_{\chi\chi^\prime}
  \Sigma_{\chi^\prime\chi}^{\text{I}(1)} P_{\chi}^{(0)}\label{Eq:current1}\\
  I^{(2)}&=&-\frac{i e}{2\hbar}
  \sum_{\chi\chi^\prime}
  \left[
    \Sigma_{\chi^\prime\chi}^{\text{I}(2)}
    P_{\chi}^{(0)}+ \Sigma_{\chi^\prime\chi}^{\text{I}(1)} P_{\chi}^{(1)}
  \right]
  \; .
  \label{Eq:current2}
\end{eqnarray}

To determine $P_{\chi}^{(0)}$ and $P_{\chi}^{(1)}$, we have to
expand the master equation, Eq.~(\ref{master}), order by order,
\begin{eqnarray}
  \label{master_1}
  0&=&\sum_{\chi} \Sigma_{\chi^\prime\chi}^{(1)} P_{\chi}^{(0)}\\
  \label{master_2}
  0&=&\sum_{\chi} \Sigma_{\chi^\prime\chi}^{(2)} P_{\chi}^{(0)}+
  \Sigma_{\chi^\prime\chi}^{(1)} P_{\chi}^{(1)} \, .
\end{eqnarray}
The evaluation of $P_{\chi}^{(0)}$ and $P_{\chi}^{(1)}$ from
Eqs.~(\ref{master_1}) and (\ref{master_2}) has to be done with
some care. As we will see below, we have to distinguish between
the two cases in which sequential tunneling is either present or
exponentially suppressed.

\subsection{Perturbation expansion in the presence of sequential tunneling}

In regime where the sequential tunneling is allowed, one can use
the perturbation expansion presented in the previous subsection.
In particular, one can determine the zeroth-order probabilities
$P_{\chi}^{(0)}$ from Eq.~(\ref{master_1}) and, then, plug the
result into Eq.~(\ref{master_2}) in order to evaluate the
first-order corrections $P_{\chi}^{(1)}$. Having calculated the
probabilities, one can use the result to get the current from
Eqs.~(\ref{Eq:current1}) and (\ref{Eq:current2}) in first and
second order, respectively.

\subsection{Perturbation expansion in the Coulomb-blockade regime}

In the Coulomb-blockade regime, several of the first-order
self-energies are exponentially small as they are associated with
energetically forbidden sequential-tunneling rates. As a
consequence, all addends in the first-order master equation,
Eq.~(\ref{master_1}), are exponentially small: either the state
$\chi$ is classically forbidden, i.e., $P_{\chi}^{(0)}$ is
exponentially suppressed, or the state $\chi$ is classically
allowed but then the corresponding self energies
$\Sigma_{\chi^\prime\chi}^{(1)}$ are exponentially small.

This is not a problem for the Coulomb-blockade valleys with an
even number of electrons, $k_{\rm B}T,\vert eV\vert \ll
\varepsilon, \varepsilon + U$ and $k_{\rm B}T, \vert eV\vert \ll -
\varepsilon, -\varepsilon - U$, since for this case, the
first-order master equation, Eq.~(\ref{master_1}), yields
$P_{\chi}^{(0)}=\delta_{\chi,0}$ and
$P_{\chi}^{(0)}=\delta_{\chi,\rm d}$, respectively, i.e., there is
only one classically-allowed dot state. The situation is different
for the Coulomb-blockade valley with an odd number of electrons,
$k_{\rm B}T,\vert eV\vert \ll -\varepsilon,\varepsilon + U$, where
both $\chi=\;\uparrow$ and $\chi=\;\downarrow$ are classically
occupied. In this case, Eq.~(\ref{master_1}) simplifies to
\begin{equation}
  \left(\begin{array}{cccc} \Sigma_{00}^{(1)} & 0 & 0 & 0 \\
    \Sigma_{\uparrow 0}^{(1)} & 0 & 0 & \Sigma_{\uparrow \rm d}^{(1)} \\
    \Sigma_{\downarrow 0}^{(1)} & 0 & 0 & \Sigma_{\downarrow \rm d}^{(1)} \\
    0 & 0 & 0 & \Sigma_{\rm dd}^{(1)}
  \end{array} \right)
  \left(\begin{array}{c}
    P_{0}^{(0)}\\
    P_{\uparrow}^{(0)}\\
    P_{\downarrow}^{(0)}\\
    P_{\rm d}^{(0)}
  \end{array} \right)= 0 \, ,
\end{equation}
i.e., we obtain $P_{0}^{(0)}=P_{\rm d}^{(0)}=0$ while the
individual occupations $P_{\uparrow}^{(0)}$ and
$P_{\downarrow}^{(0)}$ remain undetermined. Furthermore, we find
that $P_{\uparrow}^{(1)}$ and $P_{\downarrow}^{(1)}$ drop out of
the second-order master equation, Eq.~(\ref{master_2}), and the
expression for the second-order current, Eq.~(\ref{Eq:current2}),
since they are multiplied with exponentially small transition
rates $\Sigma_{\chi^\prime\chi}^{(1)}$. As a consequence, all the
needed probabilities $P_{0}^{(1)}$, $P_{\uparrow}^{(0)}$,
$P_{\downarrow}^{(0)}$, and $P_{\rm d}^{(1)}$ are determined from
Eq.~(\ref{master_2}) alone, which simplifies to
\begin{equation}
  \left(\begin{array}{cccc} \Sigma_{00}^{(1)} & \Sigma_{0\uparrow}^{(2)}
    & \Sigma_{0\downarrow}^{(2)} & 0 \\
    \Sigma_{\uparrow 0}^{(1)} & \Sigma_{\uparrow \uparrow}^{(2)} &
    \Sigma_{\uparrow \downarrow}^{(2)} & \Sigma_{\uparrow \rm d}^{(1)} \\
    \Sigma_{\downarrow 0}^{(1)} & \Sigma_{\downarrow \uparrow}^{(2)} &
    \Sigma_{\downarrow \downarrow}^{(2)} & \Sigma_{\downarrow \rm d}^{(1)} \\
    0 & \Sigma_{{\rm d} \uparrow}^{(2)} & \Sigma_{{\rm d} \downarrow}^{(2)} &
    \Sigma_{\rm dd}^{(1)}
  \end{array} \right)
  \left(\begin{array}{c}
    P_{0}^{(1)}\\
    P_{\uparrow}^{(0)}\\
    P_{\downarrow}^{(0)}\\
    P_{\rm d}^{(1)}
  \end{array} \right)= 0 \, ,
\end{equation}
plus $P_{\uparrow}^{(0)}+P_{\downarrow}^{(0)}=1$ from the
normalization condition.

If one were ignorant about the described subtlety one might
naively use the first-order master equation, Eq.~(\ref{master_1}),
with all its exponentially small (but finite) addends to obtain a
well-defined (but, in general, wrong) result for
$P_{\uparrow}^{(0)}$ and $P_{\downarrow}^{(0)}$. There are
situations, though, in which this procedure, although unjustified
by construction, leads to the correct result, namely when the
total system is symmetric under spin reversal (nonmagnetic leads,
$p=0$), or for vanishing bias voltage, $V=0$. In both cases, the
correct result $P_{\uparrow}^{(0)} = P_{\downarrow}^{(0)} = 1/2$
is ensured either by symmetry or as a consequence of detailed
balance relations. It is only for broken spin symmetry combined
with finite bias voltage $V\neq0$ that the naive procedure leads
to wrong results.

We remark that the current in the Coulomb-blockade regime far from
resonance can alternatively be calculated without the use of the
diagrammatic language. Instead one can employ a rate-equation
approach with cotunneling rates obtained in second-order
perturbation theory \cite{odintsov,nazarov,kang}. The rate
$\gamma^{\sigma' \Leftarrow \sigma}_{r'r}$ for a cotunneling
process, in which one electron leaves the dot to reservoir $r'$
and one electron enters from $r$ with the initial and final dot
state being $\sigma$ and $\sigma'$, respectively, is
\begin{eqnarray}
  \gamma^{\sigma \Leftarrow \sigma}_{r'r} &=&
  \frac{1}{2\pi}
  {\rm Re} \int d\omega [1-f(\omega-\mu_r)]f(\omega-\mu_{r'}) \times
\nonumber \\
  &&
  \left[ \frac{\Gamma^\sigma_r \Gamma^{\sigma}_{r'}}
       {(\omega-\varepsilon+i0^+)^2} +
  \frac{\Gamma^{\bar \sigma}_r \Gamma^{\bar \sigma}_{r'}}
       {(\omega-\varepsilon-U+i0^+)^2}
    \right]
\label{non-cot}
\end{eqnarray}
when the dot spin is not changed ($\sigma=\sigma'$) -- {\em
non-spin-flip cotunneling}, while we get
\begin{eqnarray}
  \gamma^{\bar \sigma \Leftarrow \sigma}_{r'r} &=&
  \frac{\Gamma^\sigma_r \Gamma^{\bar \sigma}_{r'}}{2\pi}
  {\rm Re} \int d\omega [1-f(\omega-\mu_r)]f(\omega-\mu_{r'}) \times
\nonumber \\
  &&
  \left( \frac{1}{\omega-\varepsilon+i0^+} + \frac{1}{\varepsilon+U-\omega+i0^+}
    \right)^2 \, ,
\label{sf-cot}
\end{eqnarray}
for cotunneling process in which the dot spin is flipped
($\bar\sigma$ is the opposite spin of $\sigma$) -- {\em spin-flip
cotunneling}. Here, $f(\omega-\mu_r)$ is the Fermi function of
reservoir $r$ with electro-chemical potential $\mu_r$. The
regularization $+i0^+$ is put here by hand, while it naturally
comes out within the diagrammatic formulation. There are two types
of spin-flip cotunneling processes. Each of them involves two
tunneling events, either through the same or through the two
opposite tunnel barriers. Accordingly, we refer to them as {\em
single-barrier} ($r=r'$) and {\em double-barrier cotunneling} ($r
\neq r'$). Double-barrier cotunneling contributes directly to the
current, while single-barrier cotunneling preserves the total
charge in the leads. Nevertheless, spin-flip single-barrier
cotunneling can influence the total current indirectly, by
changing of the magnetic state of the dot. The probabilities
$P_\sigma$ are obtained from the stationary rate equation
$0=\sum_{rr'} \left[ \gamma^{\downarrow \Leftarrow
\uparrow}_{r'r}P_\uparrow - \gamma^{\uparrow \Leftarrow
\downarrow}_{r'r} P_\downarrow \right] $ together with the
normalization condition $P_\uparrow + P_\downarrow =1$. The
current $I$ is, then, given by
\begin{equation}
  I = \frac{e}{\hbar} \sum_{\sigma\sigma'} \left[
  \gamma^{\sigma' \Leftarrow \sigma}_{\rm RL} -
  \gamma^{\sigma' \Leftarrow \sigma}_{\rm LR} \right] P_\sigma
  \, .
\end{equation}
This result is identical to the one obtained within the
diagrammatic technique. Close to resonance, however, it is not
sufficient to include the sequential and cotunneling processes,
but also contributions associated with renormalization of level
position, level splitting and tunnel-coupling strengths become
important. The diagrammatic language systematically takes
everything into account properly.

\subsection{Crossover scheme}

For both the case when sequential tunneling is allowed or
suppressed, we have formulated a proper perturbation expansion of
the current up to second order in the tunnel-coupling strength.
When evaluating the TMR as a function of various parameters, such
as the gate or transport voltage, one has to switch from one
scheme to the other around the threshold of sequential tunneling.
At the crossover, there is no well-defined second-order
perturbation expansion since terms of different order in $\Gamma$
are comparable in magnitude, and their ratio changes continuously
as a function of gate or transport voltage. Alternatively, we may
use a crossover scheme that smoothly crosses over from one scheme
to the other. This scheme consists of solving the master equation
with first- and second-order self energies, without expanding the
probabilities,
\begin{eqnarray}
  0&=&\sum_{\chi} \left[\Sigma_{\chi^\prime\chi}^{(1)}+
  \Sigma_{\chi^\prime\chi}^{(2)}\right]P_{\chi}\; ,
\end{eqnarray}
and plugging this into the expression for the current,
\begin{equation}
  I = -\frac{i e}{2\hbar}\sum_{\chi\chi^\prime}
  \left[ \Sigma_{\chi^\prime\chi}^{\text{I}(1)} +
    \Sigma_{\chi^\prime\chi}^{\text{I}(2)} \right]
  P_{\chi}  \, .
\end{equation}
Up to second order in $\Gamma$, this result for the current is
identical to the above-introduced accurate perturbation schemes.
Deviations are of third and higher order, which are, although
unsystematic, always small for the chosen parameters, as
otherwise, the perturbation expansion would break down anyway.

\section{Results}

\subsection{Nonmagnetic leads}

\begin{figure}[t]
  \includegraphics[width=0.87\columnwidth,height=5.2cm]{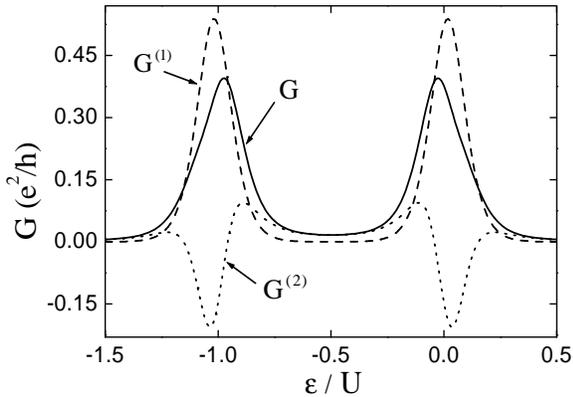}
  \caption{\label{nonmagncond} Linear conductance for nonmagnetic leads
  ($p=0$)
   as a function of
  the level position. The dashed line corresponds to the first-order
  contribution $G^{(1)}$, the dotted line represents the
  second-order conductance $G^{(2)}$ and the solid line presents the
  sum $G^{(1)}+G^{(2)}$.
  The parameters are: $k_{\rm B}T=\Gamma$ and $U=20\Gamma$.
  The figure was generated using the scheme for the perturbation expansion
  in the presence of sequential tunneling.}
\end{figure}

Before presenting the results on the TMR for quantum dots attached
to ferromagnetic leads, we illustrate the perturbation scheme
introduced above for nonmagnetic leads. In Fig.~\ref{nonmagncond}
we show the linear conductance as a function of the level position
(that can be tuned by a gate voltage), calculated to first
(dashed line) and second (dotted line) order as well as the sum of
both contributions (solid line). Resonance peaks appear when
either $\varepsilon$ or $\varepsilon+U$ crosses the Fermi energy
of the leads. Away from resonance sequential tunneling is
exponentially suppressed, and cotunneling processes dominate
transport. But also at resonance, second-order contributions are
important, as can be seen in the figure. In particular, they yield
a shift of the peak position and introduce an additional
broadening.

\subsection{Ferromagnetic leads}

We now switch to the case of ferromagnetic leads. As a consequence
of spin-dependent densities of states in the leads, the dot-lead
coupling strength becomes spin dependent as well. The coupling of
the dot level to the leads acquire a factor $(1+p)$ or $(1-p)$ for coupling
to majority or minority spins, respectively. We assume that
spin-up (spin-down) electrons in the parallel configuration
correspond to the majority (minority) electrons of the leads. In
the antiparallel configuration, on the other hand, the magnetic
moment of the right electrode is reversed, and spin-up (spin-down)
corresponds to minority (majority) electrons in the right lead.

\begin{figure}[t]
\includegraphics[width=0.75\columnwidth,height=6cm]{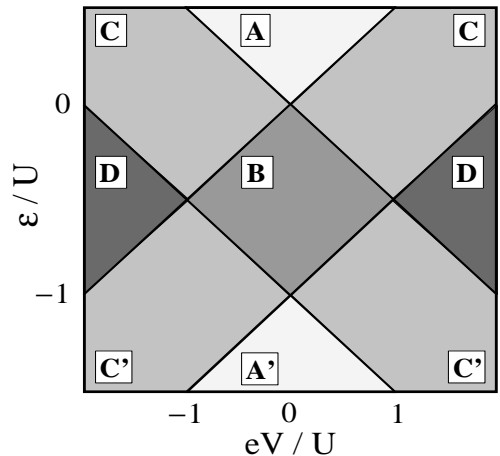}
\caption{\label{regimes}
  A sketch presenting different transport regimes.
  The respective regimes are separated by solid lines.
}
\end{figure}

One of the main results of this paper is that the TMR strongly
depends on the transport regime. The various transport regimes are
sketched in Fig.~\ref{regimes}.

In the three diamonds around $V=0$ the number of dot electrons is
fixed (to 0 in regime A, 1 in regime B, and 2 in regime A'), and
sequential tunneling is suppressed. Sequential tunneling sets in
once the bias voltage is increased above the threshold voltage,
allowing for finite occupation of two adjacent charge states (0
and 1 for regime C, and 1 and 2 for regime C'). In regime D all
charge states 0,1, and 2 are possible. By performing a
particle-hole transformation, the behavior in regime A' and C' can
be mapped to that in regime A and C, respectively.

\subsection{Sequential tunneling}

\begin{figure}[t]
  \includegraphics[width=1\columnwidth]{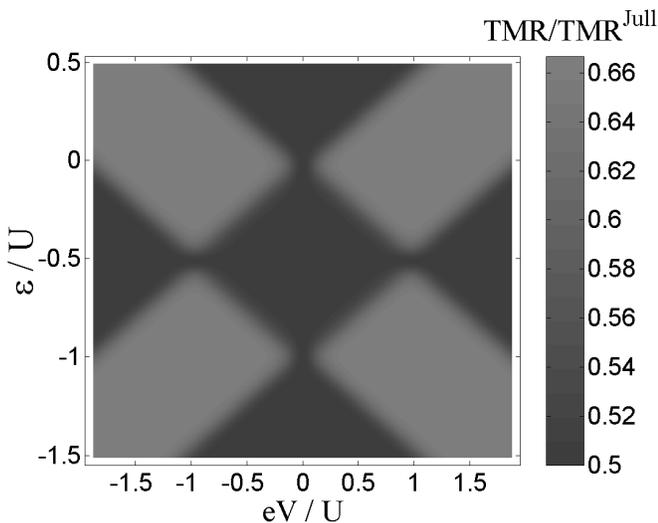}
  \caption{\label{tmrseq3d}The first-order tunnel magnetoresistance
  as a function of the bias and gate voltages. The parameters are:
  $k_{\rm B}T=1.5\Gamma$, $U=40\Gamma$, and $p=0.5$.}
\end{figure}

For reference, we list the TMR values obtained in first-order
perturbation theory (see also Fig.~\ref{tmrseq3d}). In regimes A
(and A'), B, and D, the TMR value is
\begin{equation}\label{tmrseqABD}
  {\rm TMR}_{\rm seq}^{\rm A,B,D} = \frac{p^2}{1-p^2} = \frac{1}{2}
  {\rm TMR}^{\rm Jull} \, ,
\end{equation}
while for regime C (and C') it is
\begin{equation}\label{tmrseqC}
  {\rm TMR}_{\rm seq}^{\rm C} = \frac{4p^2}{3(1-p^2)} = \frac{2}{3}
  {\rm TMR}^{\rm Jull} \, .
\end{equation}
Within sequential tunneling the TMR through a quantum-dot spin
valve is always smaller than Julliere's value for a {\em
single magnetic tunnel junction}. In the latter case, electrons
are directly tunneling from one lead to the other. The
transmission is, therefore, proportional to the product of the
(spin-dependent) densities of states of both leads, i.e.,
proportional to $(1+p)^2$ in case the spin of the transferred
electron belongs to the majority spins in both leads, $(1-p)^2$ in
case it belongs to the minority spins, and $(1+p)(1-p)$ in case it
is majority spin in one and minority spin in the other lead. The
total current for the parallel and antiparallel configurations is,
thus, proportional to $1+p^2$ and $1-p^2$, respectively, which
yields Julliere's value for the TMR.

The sequential tunneling rates in a quantum-dot spin valve involve
the (spin-dependent) density of states of one lead only and are
independent of the orientation of the other lead. To get a finite
TMR, one needs to take into account {\em nonequilibrium spin
accumulation} on the quantum dot, which is induced by the spin
dependence of the tunneling rates. In the antiparallel
configuration, the dot hosts a nonequilibrium spin accumulation
$m=(P_\uparrow - P_\downarrow)/2$ due to a different occupation of
up- and down-spin levels in the dot, $P_\uparrow \neq
P_\downarrow$. It is, thus, the spin accumulation on the dot that
mediates the information about the relative magnetic orientation
of the leads. This indirect mechanism is, however, always less
effective than a direct coupling of the two leads, which is why
the sequential-tunneling TMR is always smaller than Julliere's
value.

The result ${\rm TMR} = \frac{1}{2} {\rm TMR}^{\rm Jull}$ is
characteristic of ferromagnet/normal-metal/ferromagnet double
tunnel junctions without Coulomb interaction \cite{imamura99},
i.e., in the absence of any electron correlations, as well as for
quantum dots with vanishing interaction $U \rightarrow 0 $. For
the regime D all three charge states play a role as for
non-interacting case so the value of TMR also corresponds to this
situation. The same value is reached in the Coulomb-blockade
regimes A (A') and B, because all transport processes in this
regime are possible only due to hot electrons, which effectively
do not feel the Coulomb barrier, interaction, and correlations. In
regime C (C') Coulomb interaction is important and gives rise to
the result ${\rm TMR} = \frac{2}{3} {\rm TMR}^{\rm Jull}$. This
increased TMR is related with the presence of a nonequilibrium
spin accumulation and induced by it an additional charge
accumulation for the antiparallel alignment. To illustrate this
let us consider regime C for large bias voltages such that
electrons are always entering the dot from the left and are
leaving to the right lead. For the parallel alignment the dot
occupancy is given by
$P_{\uparrow}=P_{\downarrow}=P_{0}=\frac{1}{3} $ and $P_{\mathrm
d}=0$, while the current $I$ does not depend on the spin
polarization $p$. For the antiparallel alignment, the spin-current
conservation condition $I_{\rm L}^\sigma= I_{\rm R}^\sigma$, with
$I_r^\sigma$ being the current flowing through the barrier $r$ in
the spin channel $\sigma$, yields $(1+p)P_{0}=(1-p)P_{\uparrow}$
and $(1-p)P_{0}=(1+p)P_{\downarrow}$, i.e., the probability
$P_{0}=(1-p^2)/(3+p^2)$ to find the dot empty is reduced. Due to
the fact that the current $I \sim P_{0}$ (coming from the left
lead) for both alignments, the tunnel magnetoresistance acquires
the value $ \frac{2}{3} {\rm TMR}^{\rm Jull}$.

As in regimes A and B sequential tunneling is exponentially
suppressed, the TMR value obtained in first-order perturbation
theory is unreliable. The TMR due to cotunneling will be
significantly different, as shown below. In regimes C and D, on
the other hand, sequential tunneling is present, and second-order
corrections lead to smaller deviations only.

\subsection{Sequential tunneling plus cotunneling}

\begin{figure}[t]
  \includegraphics[width=1\columnwidth]{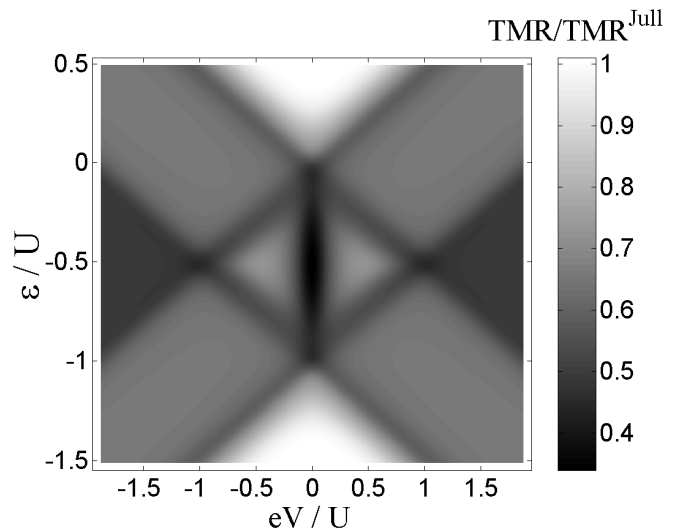}
  \caption{\label{tmr3d}The first-plus-second-order tunnel
  magnetoresistance as a function of bias and gate voltage.
  The parameters are the same as in Fig.~\ref{tmrseq3d}.
  The figure was generated using the crossover scheme.}
\end{figure}

The TMR of first- plus second-order transport is shown in
Fig.~\ref{tmr3d}, where the second-order result is obtained by the
crossover scheme. It is clear that second-order transport has the
strongest impact on the TMR in the Coulomb-blockade regime
(regimes A and B). In regime B we even find a distinctively
different behavior for the linear- and the nonlinear-response
regimes. For regimes C and D, corrections due to second-order
transport are smaller. With our theory we are able to cover all
the transport regimes including the crossover region. In the
following we analyze the various transport regimes in detail.

\subsubsection{Regime A}

In the Coulomb-blockade regime A the dot is empty, and the TMR is
just due to spin-dependent non-spin-flip cotunneling through the
dot. There is no spin accumulation on the dot. The cotunneling
rates are proportional to the product of the density of states of
the left and right leads. In this regime electrons directly tunnel
from one lead to the other similar as for a single magnetic tunnel
junction case. Thus, the current flowing in the parallel
configuration is proportional to $1+p^2$, whereas that flowing in
the antiparallel configuration is proportional to $1-p^2$. As a
consequence, the TMR is that of a single magnetic tunnel junction,
\begin{equation}
  {\rm TMR}^{\rm A} = \frac{2p^2}{1-p^2} = {\rm TMR}^{\rm Jull} \, ,
\end{equation}
i.e., twice as large as obtained within the sequential-tunneling
approximation.

In the regime A' the dot is occupied by two electrons and
transport has hole-like character with only non-spin-flip
cotunneling as for the regime A, consequently the tunnel
magnetoresistance has the same value.

\subsubsection{Regime B}

The TMR in regime B displays several nontrivial features. In
particular, it is not constant but depends on both the gate and
bias voltage. Furthermore, we find that for nonlinear response the
TMR is significantly enhanced as compared to linear response. In
contrast, the TMR in the adjacent Coulomb blockade valley with
even number of electrons, regime A, is rather trivial. This parity
effect is related to the fact that the singly-occupied dot in
regime B can be (partially) spin polarized, while the empty or
doubly-occupied dot in regime A and A' respectively is
nonmagnetic.

\begin{figure}[t]
  \includegraphics[width=0.85\columnwidth,height=10.5cm]{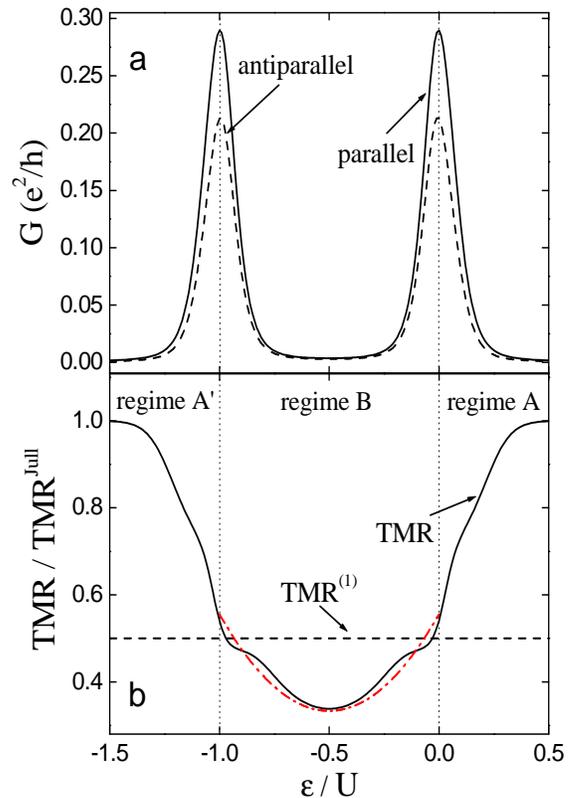}
  \caption{\label{regimeABA}The total linear conductance (a) in the
  parallel (solid line) and antiparallel (dashed line)
  configuration and the resulting tunnel magnetoresistance [solid
  line in (b)] as a function of the level position. The dashed line
  in part (b) represents the first-order tunnel magnetoresistance.
  The dotted-dashed curve presents the TMR calculated
  using the approximation Eq.~(\ref{Eq:TMRB}).
  The parameters are $k_{\rm B}T=1.5\Gamma$, $U=40\Gamma$, and $p=0.5$.
  The figure was generated using the scheme for the perturbation expansion
  in the presence of sequential tunneling.}
\end{figure}

The TMR in regime B is substantially smaller than that in regime
A. This can be understood by the fact that for a singly-occupied
dot both spin-flip and non-spin-flip cotunneling processes are
possible, in contrast to regime A and A' where only non-spin-flip
cotunneling occurs. There is a perfect symmetry in transmission
magnitude between spin-flip (non-spin-flip) processes in the
parallel and non-spin-flip (spin-flip) in the antiparallel
configuration, so in the absence of spin accumulation ($P_\uparrow
= P_\downarrow$) the resulting TMR would be reduced to zero. Only
due to the presence of spin accumulation ($P_\uparrow \neq
P_\downarrow$) for the antiparallel alignment transport is reduced
and ${\rm TMR}>0$. Therefore, the actual value of the TMR in
regime B depends in a sensitive way on the processes determining
the spin accumulation, which is a function of both the gate and
bias voltage. In particular, the different role of spin-relaxation
channels for the linear- and non-linear-response regime give rise
to qualitatively different behavior for the two cases.

\begin{figure}[t]
  \includegraphics[width=0.85\columnwidth,height=10.5cm]{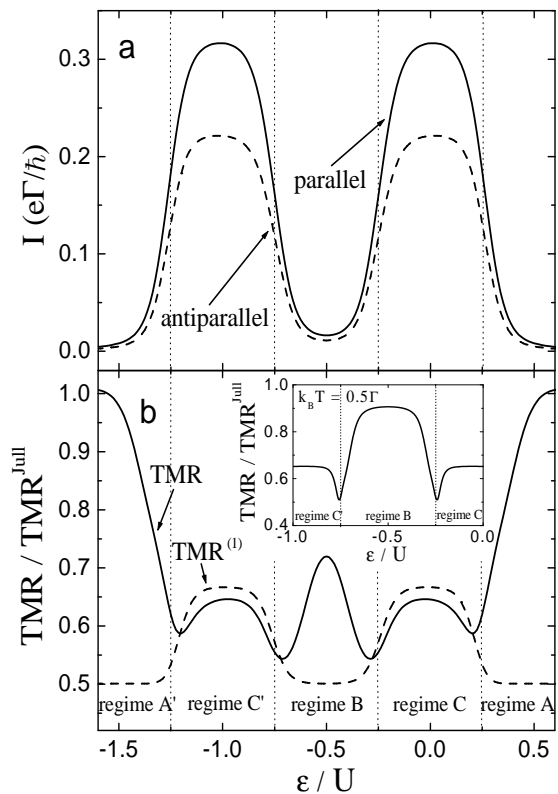}
  \caption{\label{regimeACBCA} The total currents (a) in the parallel
  (solid line) and antiparallel (dashed line) magnetic
  configurations as a function of level position for $eV=20\Gamma$.
  Part (b) shows the first-order contribution to the TMR
  (dashed line) and the total TMR (solid line).
  The inset in part (b) shows the total TMR at lower temperature,
  $k_{\rm B}T=0.5\Gamma$.
  The other parameters are the same as in Fig.~\ref{regimeABA}.
  The figure was generated using the crossover scheme.
}
\end{figure}

We first consider the {\em linear-response} TMR as a function of
level position (or gate voltage), as displayed in
Fig.~\ref{regimeABA}. The figure presents the linear conductance
in the parallel and antiparallel configurations (part a) and the
TMR (part b). We plot the first-order ${\rm TMR}^{(1)}$, which is
constant and equal to half of the Julliere's value. First of all,
one can see that the inclusion of second-order processes modifies
the TMR substantially. The total TMR is well below Julliere's
value as a consequence of spin-flip cotunneling. It is minimal in
the center of the Coulomb-blockade valley, $\varepsilon = -U/2$,
where the relative importance of spin-flip as compared to
non-spin-flip cotunneling is strongest. To estimate the
gate-voltage dependence of this relative importance we consider
the ratio of the spin-flip over the non-spin-flip cotunneling
rate, as given in Eqs.~(\ref{sf-cot}) and (\ref{non-cot}). Since
we are only interested in the gate-voltage dependence we simply
take the energy denominators at $\omega = 0$ and find that the
ratio scales with $[-1/\varepsilon + 1/(\varepsilon+U)]^2 /
[1/\varepsilon^2+1/(\varepsilon+U)^2] = 2/[1+\left( 1 +
2\varepsilon /U \right)^2 ]$, which is maximal for
$\varepsilon=-U/2$. As illustrated in Fig.~\ref{regimeABA}b, the
gate-voltage dependence of the TMR around the center is parabolic.
To obtain an approximate analytic expression for the
linear-response TMR, we specify our full result for the
Coulomb-blockade regime ($k_{\rm B}T,\;\Gamma \ll -\varepsilon, \;
\varepsilon+U$), and take into account only the lowest-order
corrections in the ratio $x/y$ with $x=\vert eV\vert,\;k_{B}T$,
$y=\vert\varepsilon\vert, \; \varepsilon+U$. To describe the
parabolic behavior, we, furthermore, expand the TMR up to
quadratic order around $\varepsilon = -U/2$ and obtain
\begin{equation}\label{Eq:TMRB}
  {\rm TMR}^{\rm B}=\frac{p^2}{1-p^2} \left[ \frac{2}{3} +
    \frac{4}{9} \left( 1 + \frac{2\varepsilon}{U} \right)^2 \right] \, .
\end{equation}
We find that the smallest TMR value is $1/3$ of that in regime A.
As seen in Fig.~\ref{regimeABA}b, this analytic expression
approximates the numerical data quite well.

We now switch to the {\em non-linear-response} regime. This case
is illustrated in Fig.~\ref{regimeACBCA}, where the currents in
the parallel and antiparallel configuration as well as the
resulting TMR are plotted as a function of the level position for
$eV=20\Gamma$. The dashed line in Fig.~\ref{regimeACBCA}b presents
the first-order TMR plotted for reference. When changing the
position of the dot level, one crosses over from regime A' over C'
to B, and then further through C to A. It can be seen that the
behavior of TMR in regime B differs significantly from that in
linear response, Fig.~\ref{regimeABA}b. Instead of a minimum, we
find a local maximum for $\varepsilon = -U/2$, as displayed in
Fig.~\ref{regimeACBCA}b. When lowering the temperature, we even
find a pronounced plateau of the TMR, with the plateau height
given by Julliere's value and the widths determined by the region
where first-order contributions are negligible. The reason for
this increased TMR is nonequilibrium spin accumulation. The
presence of double-barrier spin-flip cotunneling, on the one hand,
tends to decrease the TMR as discussed above. At the same time, on
the other hand, it gives rise to spin accumulation that increases
the TMR. As it turns out, the two effects compensate each other in
the nonlinear-response regime ($eV \gg k_{\rm B}T$), such that the
TMR equals Julliere's value as if spin-flip cotunneling were
absent. This compensation does not occur in the linear-response
regime since in that case single-barrier spin-flip cotunneling
processes become important, which do not contribute to transport
but reduce the spin accumulation. When approaching the threshold
for sequential tunneling, the TMR drops from Julliere's value to
match the first-order ${\rm TMR}^{(1)}$. At higher temperature,
such that the plateau is not yet fully developed a local maximum
still survives.

\begin{figure}[t]
\includegraphics[width=0.85\columnwidth,height=10.5cm]{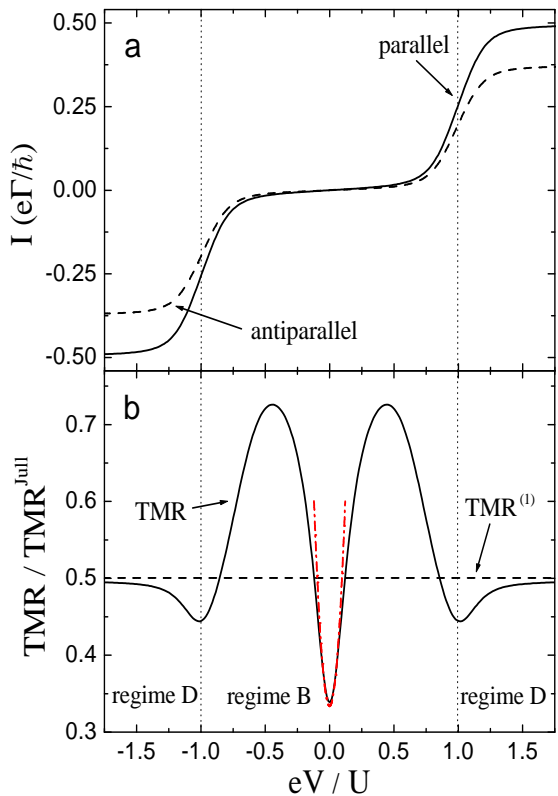}
  \caption{\label{regimeDBD}The total current (a) in the parallel
  (solid line) and antiparallel (dashed line) magnetic
  configurations as a function of the bias voltage. Part (b) shows
  the first-order contribution to the TMR (dashed line) and the total
  TMR (solid line).
  The dotted-dashed curve presents the TMR calculated
  using the approximation Eq.~(\ref{Eq:TMRBbias}).
  The parameters are: $k_{\rm B}T=1.5\Gamma$, $\varepsilon=-U/2$,
  $U=40\Gamma$, and $p=0.5$. The figure was generated using
  the crossover scheme.
}
\end{figure}

The different behavior of the linear- and nonlinear-response
regime is also nicely seen in the TMR as a function of transport
voltage. The current for the parallel and antiparallel
configuration as well as the resulting TMR is shown in
Fig.~\ref{regimeDBD} for $\varepsilon = -U/2$. Unlike the
first-order ${\rm TMR}^{(1)}$ illustrated in Fig.~\ref{regimeDBD}b
by a dashed line, the total TMR is a nonmonotonic function of the
bias voltage, which can be understood from the discussions
presented in above. For bias voltages below the threshold of
sequential tunneling, transport is dominated by cotunneling.
Double-barrier spin-flip cotunneling processes suppress the TMR as
compared to the Julliere's value. A finite spin accumulation, on
the other hand, weakens this suppression and, therefore, tends to
increase the TMR. In the linear-response regime, $|eV| \ll k_{\rm
B}T$, the presence of single-barrier spin-flip cotunneling reduces
the spin accumulation which results in a rather low TMR. This is
no longer the case at large bias, $|eV| \gg k_{\rm B}T$, where
only single-barrier spin-flip cotunneling plays no role and the
net effect of double-barrier spin-flip cotunneling on the TMR is
compensated. As a result we find an increase of the TMR in regime
B with increasing bias voltage within the limits
\begin{equation}
  \frac{1}{3}{\rm TMR}^{\rm Jull} \le {\rm TMR}^{\rm B} \le
  {\rm TMR}^{\rm Jull} \;.
\end{equation}
The minimal value is reached at $V=0$ and $\varepsilon=-U/2$, as
discussed in the previous paragraph, and the maximal value is
approached for bias voltages large as compared to temperature but
still far away from the onset of sequential tunneling. For an
approximate analytic expression of the TMR around the minimum, we
consider the symmetric Anderson model, $\varepsilon = -U/2$,
expand the TMR up to quadratic order in $|eV|/k_{\rm B}T$ and go
to the limit $|\varepsilon|\gg k_{\rm B}T$. The result,
\begin{equation}\label{Eq:TMRBbias}
  {\rm TMR}^{\rm B} = \frac{p^2}{1-p^2}\left[ \frac{2}{3}+\frac{(3-p^2)(eV)^2}
  {54(k_{\rm B}T)^2}\right] \;,
\end{equation}
which compares well with the full numerical result, as can be seen in
Fig.~\ref{regimeDBD}b.
When further increasing the bias voltage, sequential tunneling
sets in. Deep in the regime D the TMR approaches one half of
Julliere's value. As a consequence, the TMR has to decrease in the
crossover regime between regimes B and D to match the correct
asymptotic behavior, this is shown in Fig.~\ref{regimeDBD}.

There is one more extra feature directly at the threshold voltage
for sequential tunneling. At this point, sequential tunneling
dominates transport but second-order corrections are still
important. As shown in Fig.~\ref{regimeDBD}, this correction gives
rise to a local minimum of the TMR as function of the bias
voltage. To get an approximate analytic expression for the TMR at
this intersection point of regimes B, C and D, we assume
$|\varepsilon| \gg k_{\rm B}T$ and expand the TMR up to first
order in $\Gamma/(k_{\rm B}T)$ to get
\begin{eqnarray}
  \lefteqn{  {\rm TMR}^{\rm B|C|D}=\frac{p^2}{1-p^2}\times} \nonumber\\
  &&\bigg\{ 1 -\frac{\Gamma}{4\pi k_{\rm B}T}
  \left[
  \ln\left(\frac{|\varepsilon|}{\pi k_{\rm B}T} \right)
  -\Psi\left(\frac{1}{2}\right)\right] \bigg\} \;,
\end{eqnarray}
with $\Psi (x)$ being the digamma function, $\Psi(1/2)\simeq-1.96$.

\begin{figure}[t]
  \includegraphics[width=0.85\columnwidth,height=10.5cm]{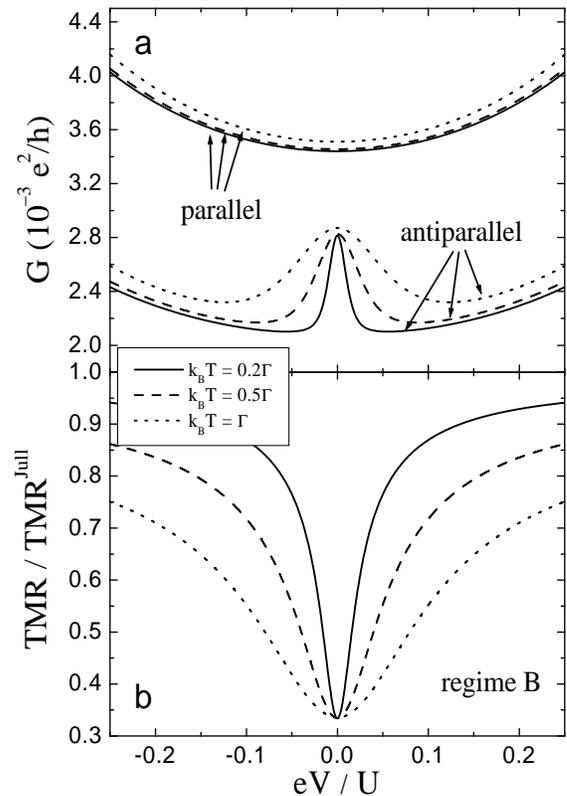}
  \caption{\label{regimeB}
  The differential conductance (a)
  for parallel and antiparallel configurations
  and the tunnel magnetoresistance (b) as a function
  of the bias voltage for different values of temperature.
  The maximum in conductance for antiparallel configuration at zero bias is clearly
  demonstrated. The other parameters are the same as in Fig.~\ref{regimeDBD}.
  Figure was generated using the scheme for
  the perturbation expansion in the Coulomb blockade regime.
  }
\end{figure}

The anomalous behavior of the TMR in the Coulomb-blockade regime
is generated by the interplay of single- and double-barrier
cotunneling for the antiparallel configuration. This is also seen
in the appearance of a pronounce zero-bias anomaly of the
differential conductance as a function of the bias voltage in the
antiparallel configuration, as we have discussed in detail in
Ref.~\onlinecite{weymann04}. For completeness we repeat here some
important facts and discuss their implications on the TMR. Deep in
the Coulomb blockade regime such that the sequential tunneling
contributions can be completely ignored, we can use the
perturbation scheme for the Coulomb blockade valley. In
Fig.~\ref{regimeB}a we show the differential conductance for both
the parallel and antiparallel configurations for different values
of the temperature. For the parallel alignment, the conductance
shows the typical cotunneling behavior, namely a smooth parabolic
dependence on the bias voltage. This contrasts with the
antiparallel configuration, for which the differential conductance
has a pronounced zero-bias peak sitting at the bottom of a
parabola. The width of the zero-bias peak is governed by
temperature, indicating different spin-accumulation behavior for
$|eV| \ll k_{\rm B}T$ and $|eV| \gg k_{\rm B}T$.

\begin{figure}[t]
\includegraphics[width=0.85\columnwidth,height=10.5cm]{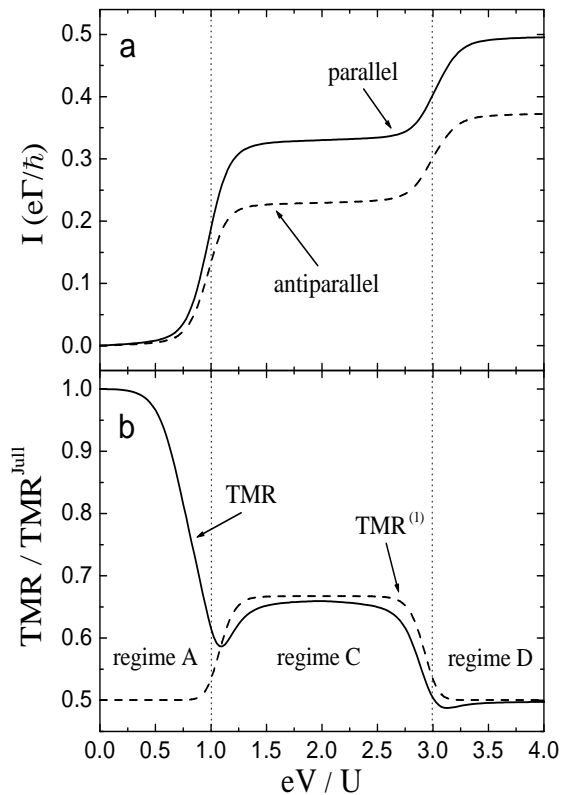}
  \caption{\label{regimeACD}The total current (a) in the parallel
  (solid line) and antiparallel (dashed line) magnetic
  configurations as a function of the bias voltage. Part (b) shows
  the first-order contribution to TMR (dashed line) and the total
  TMR (solid line).  The parameters are: $k_{\rm B}T=1.5\Gamma$,
  $\varepsilon=20\Gamma$, $U=40\Gamma$, and $p=0.5$. The figure was
  generated using the perturbation expansion in the presence of sequential
  tunneling.}
\end{figure}

\subsubsection{Regime C}

\begin{figure}[t]
  \includegraphics[width=0.85\columnwidth,height=10.5cm]{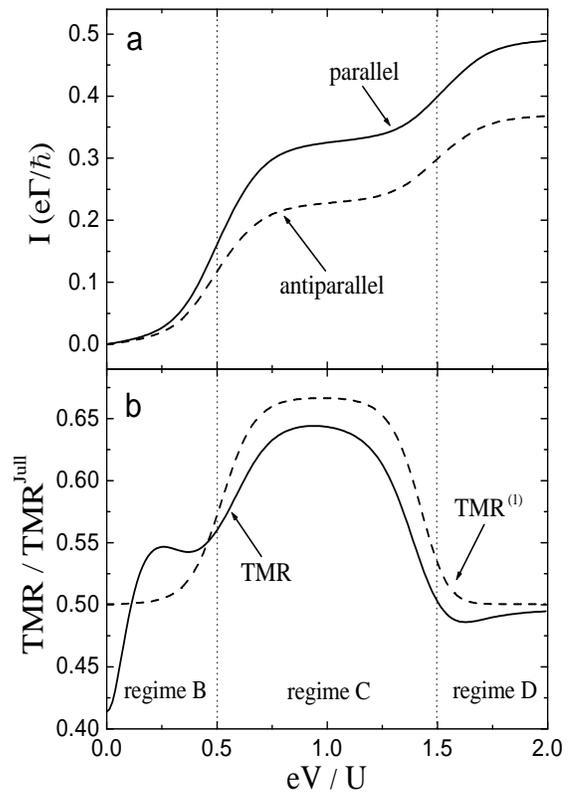}
  \caption{\label{regimeBCD}
  The total current (a) in the parallel
  (solid line) and antiparallel (dashed line) magnetic
  configuration as a function of the bias voltage. Part (b) shows
  the first-order contribution to the TMR (dashed line) and the total
  TMR (solid line).
  The parameters are: $k_{\rm B}T=1.5\Gamma$,
  $\varepsilon=-10\Gamma$, $U=40\Gamma$, and $p=0.5$. The figure was
  generated using the perturbation expansion in the presence of sequential
  tunneling.}
\end{figure}

In Fig.~\ref{regimeACD} we show the current for the parallel and
antiparallel configuration and the resulting TMR for the situation
when the dot level lies above the Fermi energy of the leads. The
first-order TMR is also shown for comparison. In this case, one
crosses over from regime A via C to D as the bias voltage is
increased. At low voltage, regime A, current is carried by
non-spin-flip cotunneling, with the TMR given by Julliere's value.
Once the threshold to regime C is reached, sequential tunneling
plays the dominant role. Second-order corrections to the current
give rise to a slightly reduced TMR as compared to the sequential
tunneling value. To find an approximate analytic expression for
this case, we consider the case of zero temperature, expand the
TMR up to first order in $\Gamma$ and assume $|\varepsilon|/U\ll
1$ to get
\begin{equation}
  {\rm TMR}^{\rm C} = \frac{p^2}{1-p^2} \left[
    \frac{4}{3} - \frac{(27+34p^2+3p^4)\Gamma}{18\pi(1-p^2) \varepsilon}
    \right] \, .
\end{equation}

At the intersection of regimes A and C the TMR develops a local
minimum. This is a consequence of the fact that when approaching
the intersection from regime C the sequential-tunneling-dominated
TMR decreases while beyond, in regime A, the TMR has to rise again
to reach Julliere's value \cite{swirkowicz02}.

In Fig.~\ref{regimeBCD} we show the current as well as the
first-order and total TMR as a function of bias voltage for
$\varepsilon=-10\Gamma$. In this case, there is a crossover from
regime B via C to D. Again, there is a local minimum of the TMR at
the threshold to sequential tunneling due to the same reason as
above.

\subsubsection{Regime D}

In regime D all the four dot states, i.e., $\chi=0,\uparrow,
\downarrow, {\rm d}$ take part in transport. This situation is
illustrated in Fig. \ref{regimeACD} for $eV>2(\varepsilon+U)$. In
this regime, transport is dominated by the first-order processes
and the influence of second-order processes is negligible.
Consequently, the value of total TMR in regime D is well described
by Eq.~(\ref{tmrseqABD}), as can be seen in Figs.~\ref{regimeDBD}b
and \ref{regimeACD}b.

\subsection{Signature of exchange field}

\begin{figure}[t]
  \includegraphics[width=0.85\columnwidth,height=10.5cm]{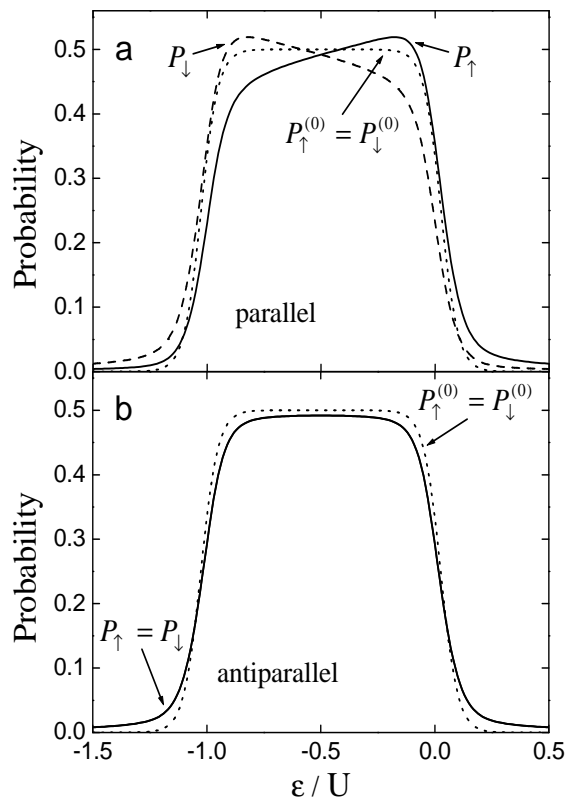}
  \caption{\label{magnlinprob}The occupation probabilities of the
  spin-up and spin-down dot levels as a function of the level
  position in the parallel (a) and antiparallel (b) configuration.
  The zeroth-order occupation probabilities for the spin-up and
  spin-down levels are equal in both magnetic configurations, and
  are represented by the dotted lines. The total occupation
  probability of the spin-up (spin-down) level is presented by the
  solid (dashed) line. In the antiparallel configuration, the dashed
  and solid lines coincide. The parameters are: $k_{\rm
  B}T=1.5\Gamma$, $U=40\Gamma$, and $p=0.5$.
  The figure was generated using the scheme for the perturbation expansion
  in the presence of sequential tunneling.}
\end{figure}

It has been predicted \cite{koenigPRL03,martinekKondo} by some of
us that the coupling of the dot levels to spin-polarized leads
gives rise to an effective exchange field seen by the quantum dot
electrons (an overview about the various effects of this exchange
field is given in Ref.~\onlinecite{koenig_springer}). This
exchange field is a consequence of both the Coulomb interaction on
the dot and the spin polarization in the leads. The contribution
coming from one lead is proportional to the degree of spin
polarization $p$ and the tunnel-coupling strength $\Gamma$. Its
direction is collinear with the leads' magnetization and its
magnitude and even the sign is a function of the level position
relative to the Fermi level. The total exchange field experienced
by the dot electrons is the (vector) sum of the two leads'
contribution. This exchange field gives rise to nontrivial
transport behavior associated with a precession of the accumulated
spin in the sequential-tunneling regime for noncollinearly
magnetized leads \cite{koenigPRL03,braun,braig04} and leads to a
splitting of the Kondo resonance in the strong-coupling limit
\cite{martinekKondo,choi}, as experimentally observed recently
\cite{pasupathy}. By applying our diagrammatic technique, the
exchange field is automatically included.

As we argue in the following, the exchange field will, under
certain circumstances, also show up in the parameter regime
studied in this paper, namely as an {\it equilibrium} spin
polarization of the dot. This is distinctively different from the
{\it nonequilibrium} spin accumulation discussed in the previous
sections. The latter is a nonequilibrium effect that changes sign
with bias reversal and, in particular, vanishes for zero bias
voltage. In contrast, a finite spin polarization at equilibrium
can only occur when the dot level is spin split by either an
external magnetic field or by the intrinsic exchange field that we
want to address now.

In the antiparallel configuration, and for symmetric coupling to
and equal spin polarization of the leads, the exchange-field
contributions from the two leads exactly cancel out each other
since they are of equal magnitude but pointing in opposite
directions. This is different for the parallel configuration, for
which the contributions from the two leads add up to some finite
value.

To lowest (zeroth) order in the tunnel coupling strengths
$\Gamma$, the equilibrium probabilities for occupation with spin
$\sigma=\uparrow,\downarrow$ are determined by the Boltzmann
factors $P_{\uparrow}^{(0)} = P_{\downarrow}^{(0)} =
\exp(-\beta\varepsilon)/Z$, where $Z$ denotes the partition
function. Since the exchange field is proportional to $\Gamma$, it
does not affect the zeroth-order occupation probabilities, i.e.,
the sequential-tunneling approximation is not able to describe the
exchange-field induced spin polarization. This is shown in
Fig.~\ref{magnlinprob}, where the equilibrium probabilities
calculated to zeroth- and zeroth- plus first-order in the dot-lead
coupling are presented. A finite spin polarization for the
parallel configuration is only generated by the first-order
corrections $P_{\uparrow}^{(1)} \neq P_{\downarrow}^{(1)}$, that
we obtain by solving the master equation given by
Eq.~(\ref{master_2}). The $\varepsilon$-dependence of the spin
polarization seen in Fig.~\ref{magnlinprob} reflects the
$\varepsilon$-dependence of the exchange field. The exchange field
for a particle-hole symmetric band vanishes in the middle of the
Coulomb blockade valley, $\varepsilon=-U/2$, and has different
sign on either side. As a consequence the dot polarization changes
sign as well.

Since in regime B $\Sigma_{\chi \sigma}^{\rm I(1)}$ are
exponentially suppressed, the exchange splitting and probabilities
$P_{\sigma}^{(1)}$ do not affect the second-order transport. These
probabilities affect only higher-order transport contributions,
which at low temperature $T \lesssim T_{\rm K}$ lead to the Kondo
effect \cite{martinekKondo,choi,pasupathy}.

\section{Summary}

We have discussed electronic transport through quantum dots
coupled to ferromagnetic leads. Based on a formalism that allows
for a systematic perturbation expansion in the tunnel coupling
strength, we analyzed the TMR through a single-level quantum dot
for the linear- and nonlinear-response regime, at or off resonance,
with an even or odd dot electron number. We found different TMR
values for different transport regimes. In addition to the full
numerical results we provided approximate analytic expressions for
various limiting cases.
The most important findings are:\\
(i) Except for the Coulomb-blockade valley with an even
dot-electron number and the nonlinear-response regime of the Coulomb-blockade
valley with an odd dot-electron number, the TMR is below that of a single
magnetic tunnel junction.\\
(ii) There is an even-odd asymmetry between the Coulomb-blockade
valleys with an even or odd number of electrons, that is related
to the absence or presence of spin-flip cotunneling, respectively.\\
(iii) In the Coulomb-blockade valley with an odd number of
electrons, the TMR values for the linear and nonlinear response
regimes differ strongly from each other, associated with different
spin-relaxation processes that affect the spin accumulation.\\
(iv) The linear-response TMR in the Coulomb-blockade valley with
an odd number of electrons is a function of gate voltage, which
reflects the relative
importance of spin-flip and non-spin-flip cotunneling.\\
(v) The TMR at the onset of sequential tunneling displays a local minimum,
which is a consequence of interpolating the TMR behavior away from resonance.

\begin{acknowledgments}

We thank Matthias Braun, Matthias Hettler, Ken Imura, Sadamichi
Maekawa, Axel Thielmann for helpful discussions. The work was
supported by the Deutsche Forschungsgemeinschaft through SFB491
and GRK726, the Polish State Committee for Scientific Research
through the projects PBZ/KBN/044/P03/2001 and 2 P03B 116 25,
"Spintronics" RT Network of the EC Grant No. RTN2-2001-00440, and
the Center of Excellence for Magnetic and Molecular Materials for
Future Electronics within the EC Contract G5MA-CT-2002-04049.

\end{acknowledgments}


\appendix

\section{Diagrammatic technique}

In this Appendix we present general rules in energy space for
calculating contributions of various diagrams. We also present an
exemplary calculation of one of the second-order self-energies.
Afterwards, we show how to determine self-energies contributing to
electric current.

\subsection{Rules in energy space}

Contribution of a particular diagram to the self-energy
$\Sigma_{\chi'\chi}$ can be found following the general rules in
the energy space:
\begin{enumerate}
\item Draw all topologically different diagrams with fixed time
ordering and position of vertices. Connect the vertices by
tunneling lines. Assign the energies of respective quantum dot
states to the forward and backward propagators. To each tunneling
line assign a frequency $\omega$, the spin of tunneling electron
and label of the junction.
\item Tunneling lines acquire arrows indicating whether an
electron leaves or enters the dot. For tunneling lines going
forward with respect to the Keldysh contour assign a factor
$\gamma^{-\sigma}_{r}(\omega)$, whereas for tunneling lines going
backward assign $\gamma^{+\sigma}_{r}(\omega)$.
\item For each time interval on the real axis limited by two
adjacent vertices draw a vertical line inside the interval and
assign a resolvent $1/(\Delta E+i0^+)$, with $\Delta E$ being the
difference of all energies crossing the vertical line from right
minus all energies crossing the vertical line from left.
\item Each diagram gets a prefactor $(-1)^{b+c}$, with $b$ being
the number of vertices lying on the backward propagator and $c$
denoting the number of crossings of the tunneling lines.
\item Each internal vertex represents a matrix element $\bra{\chi}
A \ket{\chi'}$, with $A$ being a dot operator,
$A=d_{\sigma}^{\dagger},d_{\sigma}$. Consequently, a minus sign
may appear due to these matrix elements. This is because $\ket{\rm
d}=d_{\sigma}^{\dagger}\ket{\bar{\sigma}}=
-d_{\bar{\sigma}}^{\dagger}\ket{\sigma}$ (depending on the
definition of state $\ket{\rm d}$), where $\sigma=\uparrow$ or
$\sigma=\downarrow$. To account for this factor, multiply each
diagram by $(-1)^{m}$, where $m$ is the number of vertices
connecting the spin-$\sigma$ state with doubly occupied state.
\item Integrate over all frequencies and sum up over the
reservoirs.
\end{enumerate}
The parameters $\gamma^{\pm\sigma}_{r}(\omega)$ are defined as
\begin{eqnarray}
  \gamma^{+\sigma}_{r}(\omega)&=&
  \frac{\Gamma^{\sigma}_{r}}{2\pi}f(\omega-\mu_{r}),\\
  \gamma^{-\sigma}_{r}(\omega)&=&
  \frac{\Gamma^{\sigma}_{r}}{2\pi}[1-f(\omega-\mu_{r})],
\end{eqnarray}
with $f(x)$ being the Fermi-Dirac distribution function,
$f(x)=1/\left[\exp(x/k_{\rm B}T)+1\right]$, and $\mu_{r}$
representing the electrochemical potential of lead $r$.

\subsection{Calculation of $\Sigma_{\bar{\sigma}\sigma}^{(2)}$}

In order to find the zeroth-order and first-order probabilities,
one needs to determine all the self-energies of first and second
order in $\Gamma$. Below, we present an exemplary calculation of
one of the second-order self-energies,
$\Sigma_{\bar{\sigma}\sigma}^{(2)}$. The equation for
$\Sigma_{\bar{\sigma}\sigma}^{(2)}$ can be graphically presented
as
\begin{widetext}
  \begin{equation}\label{Eq:diagrams}
    \includegraphics[width=0.75\columnwidth]{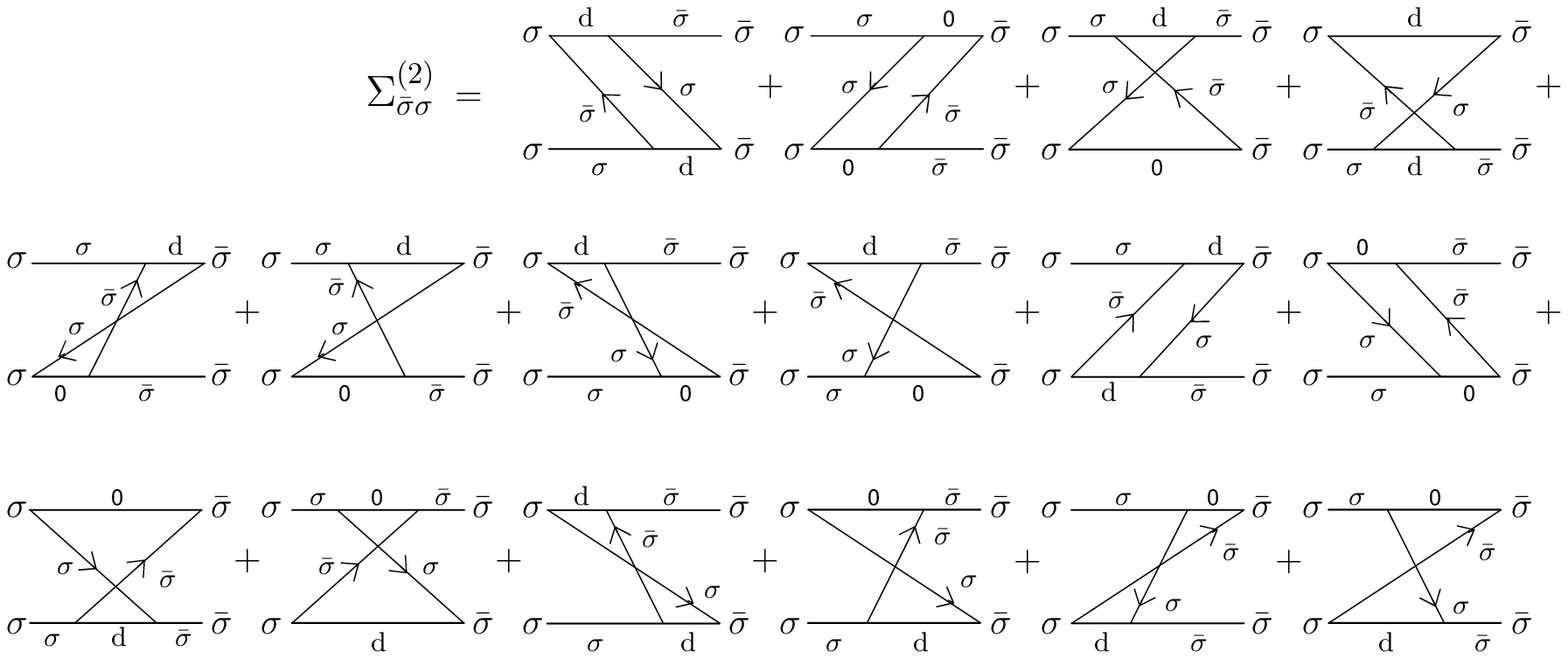}
  \end{equation}
\end{widetext}
To calculate the self-energy, it is necessary to evaluate each
contributing diagram. As an example, we present calculation of the
third diagram of  Eq. (\ref{Eq:diagrams}). Following the general
rules described above, the corresponding contribution,
$\Lambda_{3}$, is given by
\begin{widetext}
  \begin{eqnarray}\label{Eq:diagram}
    \Lambda_{3}=(-1)^{2+1}(-1)^{1}\sum_{r_{1},r_{2}}\iint d\omega
    _1 d\omega _2 \gamma_{r_1}^{-\sigma}(\omega_{1})\gamma_{r_2}
    ^{+\bar{\sigma}}(\omega_{2})\frac{1}{\omega_{1}
    -\varepsilon_{\sigma}+i0^+} \frac{1}{\omega_{1}+\omega_{2}
    -\varepsilon_{\sigma}-\varepsilon_{\bar{\sigma}}-U+i0^+} \frac{1}{\omega_{2}
    -\varepsilon_{\bar{\sigma}}+i0^+}\,.
  \end{eqnarray}
\end{widetext}
The first (second) factor on the right-hand side follows from the
rule 4 (5). There are also three resolvents according to the rule
(3). Among the various diagrams contributing to
$\Sigma_{\bar{\sigma}\sigma}^{(2)}$, there is a diagram (eleventh
in Eq. \ref{Eq:diagrams}) whose contribution is equal to minus
complex conjugate of the contribution due to the third diagram,
$\Lambda_{11}=-{\rm Re} (\Lambda_{3})+i{\rm Im} (\Lambda_{3})$.
This can be shown by interchanging the backward and forward
propagators and changing the direction of the tunneling lines. As
a consequence, the real parts of these diagrams cancel, whereas
the imaginary parts add to each other. Thus, it is necessary to
determine only the imaginary part of one of those two diagrams,
$\Lambda_{3}+\Lambda_{11}=2i{\rm Im} (\Lambda_{3})$. After contour
integration, the imaginary part of $\Lambda_{3}$ is given by
\begin{eqnarray}
  \lefteqn{{\rm Im} (\Lambda_{3})=\frac{\pi}{U}\sum_{r_{1},r_{2}}\bigg[
  \gamma_{r_{1}}^{-\sigma}(\varepsilon_{\sigma})
  A_{1r_{2}}^{+\bar{\sigma}}(\varepsilon_{\bar{\sigma}})+
  \gamma_{r_{2}}^{+\bar{\sigma}}(\varepsilon_{\bar{\sigma}})
  A_{1r_{1}}^{-\sigma}(\varepsilon_{\sigma}) }\nonumber\\
  && - \frac{\Gamma_{r_{1}}^{\sigma}}{2\pi}
  f_{\rm B}(\mu_{r_{1}}+\mu_{r_{2}}-\varepsilon_{\sigma}-\varepsilon_{\bar{\sigma}}-U)
  X_{1r_{2}}^{+\bar{\sigma}}
  (2\mu_{r_{2}}-\varepsilon_{\bar{\sigma}}-U)\nonumber\\
  &&-\frac{\Gamma_{r_{2}}^{\bar{\sigma}}}{2\pi}
  f_{\rm B}(\varepsilon_{\sigma}+\varepsilon_{\bar{\sigma}}+U-\mu_{r_{1}}-\mu_{r_{2}})
  X_{1r_{1}}^{+\sigma}(\varepsilon_{\sigma}) \bigg],
\end{eqnarray}
with $f_{\rm B}(x)$ being the Bose-Einstein distribution function
$f_{\rm B}(x)=1/\left[\exp(x/k_{\rm B}T)-1\right]$. The
corresponding coefficients $A_{\alpha
r}^{\pm\sigma}(\varepsilon_{\sigma})$ are defined as, $A_{\alpha
r}^{\pm\sigma}(\varepsilon_{\sigma})= X_{\alpha
r}^{\pm\sigma}(\varepsilon_{\sigma})- X_{\alpha
r}^{\pm\sigma}(\varepsilon_{\sigma}+U)$, with $X_{\alpha
r}^{\pm\sigma}(\varepsilon_\sigma)=\pm \Gamma_{r}^{\sigma}/(2\pi)
B_\alpha(\varepsilon_\sigma -\mu_r)$ and $B_\alpha(x)$ given by
\begin{equation*}
  B_{\alpha+1}(x)=\frac{d^{(\alpha)}}{dx^{(\alpha)}}
  {\rm Re} \bigg[ \Psi\left(
  \frac{1}{2}+i\frac{x}{2\pi k_{\rm B}T}\right)-\ln\left(
  \frac{\rm W}{2 \pi k_{\rm B}T}\right)\bigg],
\end{equation*}
where $\Psi(z)$ is the digamma function, and we have used the
Lorentzian cutoff function of the form $\rho_{\nu}(\omega)= {\rm
W}^2/[(\omega-\mu_{\nu})^2+{\rm W}^2]$, with ${\rm W}$ being the
cutoff parameter. As contribution from a single diagram may depend
on ${\rm W}$, the final result does not. In the calculations the
cutoff parameter was taken to be equal to $100\Gamma$.

In a similar way, one can calculate contributions of all diagrams,
which give
\begin{widetext}
  \begin{eqnarray}
    \Sigma_{\bar{\sigma}\sigma}^{(2)}&=&-2\pi i \sum_{r_1, r_2}
    \Bigg\{
      \gamma_{r_1}^{-\sigma}(\varepsilon_{\sigma})
      X_{2r_2}^{+\bar{\sigma}}(\varepsilon_{\bar{\sigma}})+
      \gamma_{r_1}^{+\bar{\sigma}}(\varepsilon_{\bar{\sigma}})
      X_{2r_2}^{-\sigma}(\varepsilon_{\sigma})\nonumber\\
      &&+ \gamma_{r_1}^{-\sigma}(\varepsilon_{\sigma}+U)
      X_{2r_2}^{+\bar{\sigma}}(\varepsilon_{\bar{\sigma}}+U)+
      \gamma_{r_1}^{+\bar{\sigma}}(\varepsilon_{\bar{\sigma}}+U)
      X_{2r_2}^{-\sigma}(\varepsilon_{\sigma}+U)
      \nonumber\\
      &&-f_{\rm B}(\mu_{r_{1}}-\mu_{r_{2}}+
      \varepsilon_{\bar{\sigma}}-\varepsilon_{\sigma})
      \left\{\frac{\Gamma_{r_{2}}^{\bar{\sigma}}}{2\pi}
        \left[X_{2r_1}^{+\sigma}(\varepsilon_{\sigma})+
          X_{2r_1}^{+\sigma}(\varepsilon_{\sigma}+U)
          +\frac{2}{U}A_{r_1}^{+\sigma}(\varepsilon_{\sigma})
        \right]
      \right.\nonumber\\
    &&\left.- \frac{\Gamma_{r_{1}}^{\sigma}}{2\pi}
        \left[
          X_{2r_2}^{+\bar{\sigma}}(\varepsilon_{\bar{\sigma}})+
          X_{2r_2}^{+\bar{\sigma}}(\varepsilon_{\bar{\sigma}}+U)
          +\frac{2}{U}A_{r_2}^{+\bar{\sigma}}(\varepsilon_{\bar{\sigma}})
        \right]
      \right\}
    \Bigg\}
  \end{eqnarray}
\end{widetext}

\subsection{Diagrams contributing to the current}

To find current flowing through the system, one has to determine
the self-energies $\Sigma^{\rm I}$, see  Eq. (\ref{Eq:current1})
or (\ref{Eq:current2}). This can be done by realizing that each
term of the expansion of the current operator $\hat{I}$ is equal
to the corresponding expansion term of the reduced density matrix
multiplied by a factor of $e/\hbar$. The only difference is that
now for each external vertex lying on the upper (lower) branch of
the Keldysh contour,  corresponding to tunneling of an electron
into the left (right) or out of the right (left) lead, we have a
multiplicative factor +1/2, whereas for each external vertex on
the upper (lower) branch of the contour, describing tunneling of
an electron into the right (left) or out of the left (right) lead,
there is a factor of -1/2.

We have determined all the first-order and second-order
self-energies contributing to electrical current, $\Sigma^{{\rm
I}(1)}$ and $\Sigma^{{\rm I}(2)}$, and found that from the
first-order self-energies only $\Sigma_{0 \sigma}^{{\rm I}(1)}$,
$\Sigma_{\sigma 0}^{{\rm I}(1)}$, $\Sigma_{\sigma {\rm d}}^{{\rm
I}(1)}$, $\Sigma_{{\rm d} \sigma}^{{\rm I}(1)}$ give nonzero
contributions. In the case of the second-order self-energies we
found $\Sigma_{\chi\chi}^{{\rm I}(2)}=0$, with
$\chi=0,\uparrow,\downarrow,{\rm d}$. This is however only the
case for the current operator defined as $\hat{I}=(\hat{I}_{\rm
R}-\hat{I}_{\rm L})/2$, where $\hat{I}_{r}$ is the current
operator for electrons tunneling to the lead $r$.


\end{document}